\documentclass{aa}
\usepackage{graphicx}
\usepackage{color}
\usepackage{txfonts}
\usepackage{lscape}
\begin{document}

   \authorrunning{Li \& Zhang}

   \title{Subarcsecond Bright Points and Quasi-periodic Upflows
Below a Quiescent Filament Observed by the IRIS}

   \author{T. Li, J. Zhang}

   \institute{Key Laboratory of Solar Activity, National Astronomical
   Observatories, Chinese Academy of Sciences, 100012 Beijing, China\\
              \email{liting@nao.cas.cn}}
   \date{Received ????; accepted ????}


  \abstract
   {The new Interface Region
Imaging Spectrograph (IRIS) mission provides high-resolution
observations of UV spectra and slit-jaw images (SJIs). These data
has become available for investigating the dynamic features in the
transition region (TR) below the on-disk filaments.}
   {The driver of ``counter-streaming" flows along the filament spine is
still unknown yet. The magnetic structures and the upflows at the
footpoints of the filaments, and their relations with the filament
mainbody have not been well understood. We study the dynamic
evolution at the footpoints of filaments in order to find some clues
for these unsolved questions.}
   {Using UV spectra and SJIs from the IRIS, and  coronal images and
magnetograms from the Solar Dynamics Observatory (SDO), we present
the new features in a quiescent filament channel: subarcsecond
bright points (BPs) and quasi-periodic upflows.}
   {The BPs in the TR have a spatial scale of about 350$-$580 km and
lifetime of more than several tens of minutes. They are located at
stronger magnetic structures in the filament channel, with magnetic
flux of about 10$^{17}$$-$10$^{18}$ Mx. Quasi-periodic brightenings
and upflows are observed in the BPs and the period is about 4$-$5
min. The BP and the associated jet-like upflow comprise a
``tadpole-shaped" structure. The upflows move along bright filament
threads and their directions are almost parallel to the spine of the
filament. The upflows initiated from the BPs with opposite polarity
magnetic fields have opposite directions. The velocity of the
upflows in plane of sky is about 5$-$50 km s$^{-1}$. The emission
line of Si IV 1402.77 {\AA} at the locations of upflows exhibits
obvious blueshifts of about 5$-$30 km s$^{-1}$, and the line profile
is broadened with the width of more than 20 km s$^{-1}$.}
   {The BPs seem to be the bases of filament threads and the upflows
are able to convey mass for the dynamic balance of the filament. The
``counter-streaming" flows in previous observations may be caused by
the propagation of bi-directional upflows initiated from opposite
polarity magnetic fields. We suggest that quasi-periodic
brightenings of BPs and quasi-periodic upflows result from
small-scale oscillatory magnetic reconnections, which are modulated
by solar p-mode waves.}

   \keywords{Sun: filaments, prominences--Sun: transition region -- Sun: UV radiation}

   \maketitle
%

\section{Introduction}

Quiescent filaments are large structures of cool and dense plasma
embedded in the surrounding hotter corona of quiet regions. Despite
over a century of observations, a big question still remains: how
the filament plasma forms and evolves. It is recognized that the
filament material should come from the chromosphere, rather than the
rarefied corona (Pikel'Ner 1971; Saito \& Tandberg-Hanssen 1973).
Several models have been proposed to explain the formation process
of filament material. In injection models, chromospheric plasma is
propelled directly into the corona through the reconnection of
magnetic field during canceling flux (Wang 1999; Chae et al. 2001;
Litvinenko et al. 2007). Levitation models propose that cool plasma
is lifted by flux rope emergence from below the photosphere or
post-reconnection relaxation associated with flux cancellation (van
Ballegooijen \& Martens 1989; Priest et al. 1996). In
evaporation-condensation models, chromospheric plasma is evaporated
due to thermal non-equilibrium and ultimately condensed in the
corona to form cool filament threads (Karpen et al. 2005; Xia et al.
2012; Liu et al. 2012; Berger et al. 2012). The dynamic evolution at
the footpoints of the filaments could provide important information
for our understanding of the formation mechanism of the filaments
(Parenti 2014; Li et al. 2015).


Bright points (BPs) are ubiquitous small-scale bright dynamical
features at various atmosphere layers. BPs are frequently observed
in X-rays and extreme ultraviolet (EUV) wavelengths and correspond
to coronal bright points (CBPs). The spatial sizes of CBPs are
around 5$-$20$\arcsec$ and their lifetime ranges from several hours
to 2 days (Vaiana et al. 1973; Golub et al. 1974; Li et al. 2013).
CBPs are often located at network magnetic features of opposite
polarity, with the total fluxes of about 10$^{19}$$-$10$^{20}$ Mx
(Golub et al. 1976; Hong et al. 2014). The observations showed that
some CBPs are associated with quasi-periodic impulsive flashes, with
periods of a few minutes to hours (Habbal et al. 1990; Ning \& Guo
2014; Samanta et al. 2015). Some authors suggest that small-scale
repeated magnetic reconnections result in the quasi-periodic
variation in the intensity of CBPs (Madjarska et al. 2003; Doyle et
al. 2006; Zhang et al. 2012). Recently, R{\'e}gnier et al. (2014)
reported the existence of subarcsecond EUV bright dots at the edge
of the active region with a characteristic length of 680 km from
observations of the High-resolution Coronal (Hi-C) instrument. The
bright dots are short-lived (25 s) and appear at the footpoints of
large-scale coronal loops. Tian et al. (2014a) found similar
subarcsecond bright dots above sunspots in the transition region
(TR) with the observations of the Interface Region Imaging
Spectrograph (IRIS; De Pontieu et al. 2014).

The majority of previous observations about filaments were taken in
H$\alpha$ and EUV wavelengths, and the dynamical features below the
on$-$disk filaments could not be seen as a result of the abundant
Lyman continuum absorption and the ``volume$-$blocking" effect (Wang
et al. 1998; Heinzel et al. 2001). The recently launched IRIS
mission is now providing high-resolution observations of the TR and
the chromosphere. The slit-jaw images (SJIs) from the IRIS are
centered at 1330, 1400, 2796 and 2832 {\AA} (De Pontieu et al.
2014). The filaments seem semitransparent and optically thin in the
channels of 1330 and 1400 {\AA} and the dynamic evolution at the
footpoints of the filaments could be seen. Thus in this work we
investigate the small-scale activities at the footpoints of the
filament in the TR  by combining imaging and spectroscopic
observations of the IRIS and try to find some clues to the material
source of the filaments. Here, new dynamical features in the TR
below the quiescent filaments are found: subarcsecond BPs and
quasi-periodic upflows. We investigate the physical properties of
BPs, the quasi-periodic patterns of BPs and associated upflows.


\begin{figure}
   \centering
   \includegraphics[bb=46 65 509 752,width=8.8cm]{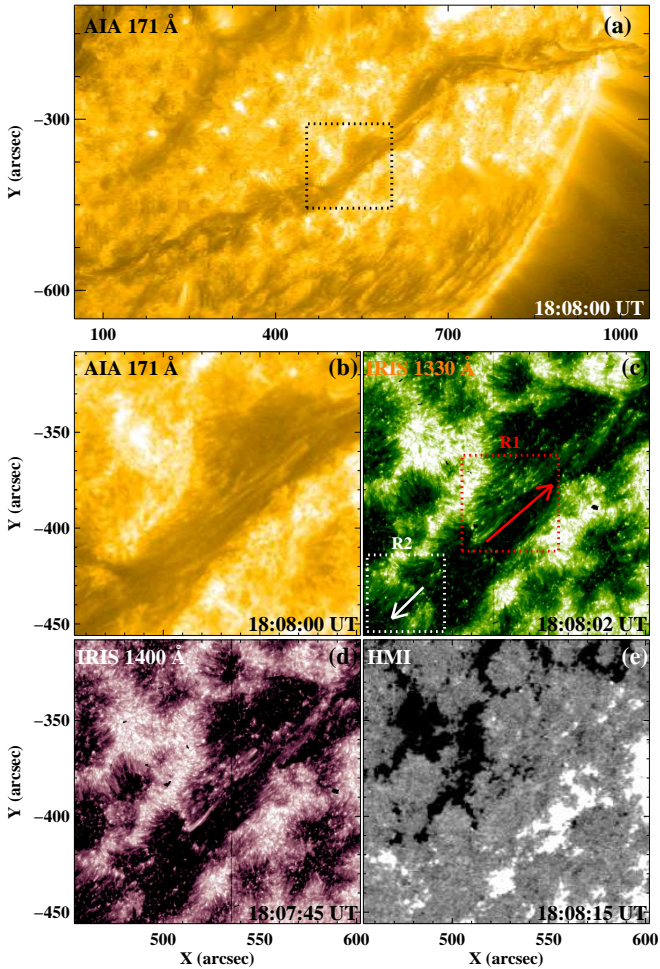}
   \caption{Panel (a): SDO/AIA 171 {\AA} image showing the large
quiescent filament on 2015 Feb 13. The black square displays the FOV
of the zoomed images in panels (b)-(e). Panels (b)-(e):
multi-wavelength images of SDO/AIA, IRIS 1330 {\AA}, IRIS 1400 {\AA}
and SDO/HMI LOS magnetogram showing the subarcsecond BPs underlying
the filament and bi-directional (red and white arrows) loop-like
structures at regions ``R1" (red square) and ``R2" (white square).}
   \label{F:lineprofiles}%
\end{figure}


\section{Observations and Data Analysis}\label{S:obs}

The IRIS observations were taken from 18:01 UT to 18:45 UT on 2015
February 13, with the field of view (FOV) covering part of a
quiescent filament in the southern hemisphere. The SJIs centered at
1330, 1400 and 2796 {\AA} are used to analyze the evolution of BPs,
with 0$\arcsec$.33$-$0$\arcsec$.4 spatial resolution. The
observations have a long exposure time of 15 s and a temporal
cadence of $\sim$ 66 s. The spectral data are taken in a 8$-$step
raster mode with a step cadence of 16.4 s and a spectral dispersion
of $\sim$0.025 {\AA} pixel$^{-1}$. The SJIs at 1330 {\AA} and
spectral data of the Si IV 1402.77 {\AA} line are mainly analyzed.
The C II 1335.71 {\AA} line is formed in the lower TR and
corresponds to a temperature of about 2.5$\times$10$^{4}$ K (Li et
al. 2014). The Si IV 1402.77 {\AA} line is formed in the middle TR
with a temperature of about 8.0$\times$10$^{4}$ K. The calibrated
level 2 data were used in our study. The EUV observations from the
Atmospheric Imaging Assembly (AIA; Lemen et al. 2012) onboard the
Solar Dynamics Observatory (SDO; Pesnell et al. 2012) are used to
compare with the IRIS data. Cross-correlation between the AIA 1600
{\AA} image and the IRIS SJI 1400 {\AA} image was used for the
coalignment (Tian et al. 2014b). The full-disk line-of-sight
magnetograms from the Helioseismic and Magnetic Imager (HMI;
Scherrer et al. 2012) are also applied to analyze the magnetic
structures at the BPs.

\begin{figure*}
   \centering
   \sidecaption
   \includegraphics[bb=48 255 509 564,width=12cm]{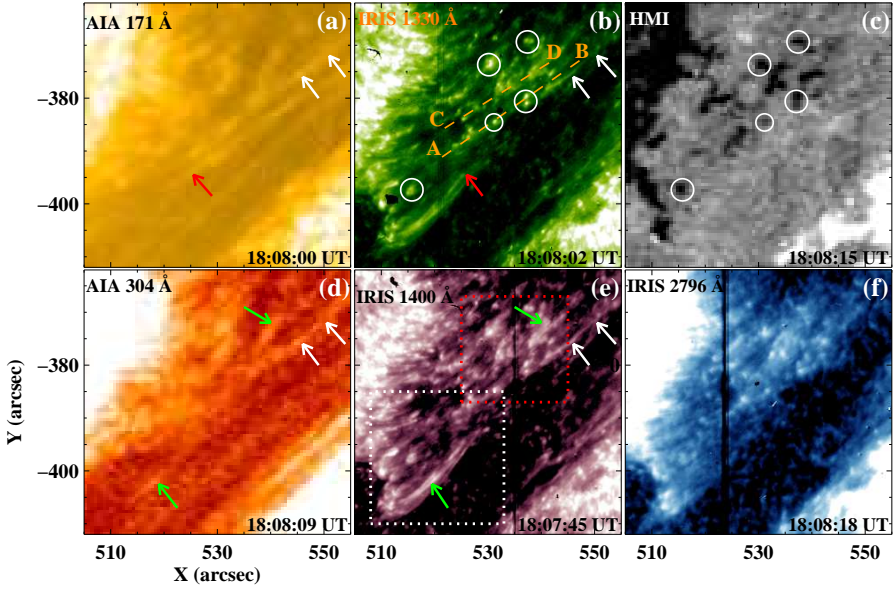}
   \caption{Appearance of BPs and loop-like structures at region ``R1".
The white circles outline the locations of several BPs in the 1330
{\AA} image and HMI LOS magnetogram. The arrows in panels (a)-(b)
and (d)-(e) point to the loop-like structures in close proximity to
some BPs. Dashed lines ``A$-$B" and ``C$-$D" (panel b) show the cut
positions used to obtain the stack plots shown in Figures 5(a)-(b).
Red and white squares in panel (e) respectively denote the FOVs of
Figures 3(a)-(d) and 3(e)-(l).
   }
   \label{F:lineprofiles}%
\end{figure*}

\section{Results}

The large-scale quiescent filament stretches across the solar
southern hemisphere, with a length of $\sim$1000 Mm (Figure 1a). The
middle part of the filament is covered by the FOV of the IRIS
(148$\arcsec$$\times$148$\arcsec$). In the zoomed 171 {\AA} image
(Figure 1b), several bright thread-like structures were observed
along the spine of the filament. At the bases of these threads, a
few grain-like knots could be obscurely seen. High-resolution
observations of 1330 and 1400 {\AA} clearly showed several tens of
small-scale BPs underlying the filament (Figures 1c-d). These BPs
are mainly located at two areas below the filament (regions ``R1"
and ``R2" in Figure 1c). Most of the BPs in regions ``R1" and ``R2"
are respectively located at opposite polarity magnetic fields
(Figure 1e). We have selected 22 relatively isolated BPs in the two
regions and obtain the intensity-location profiles along the slices
that pass through the centers of the BPs. Then the Gaussian function
is used to fit the profiles and the full width at half maximum
(FWHW) of the Gaussian fitting profile is thought to be the spatial
size of each BP. The average size of these BPs is 440 km (about 4
pixels), with the maximum value of 580 km and the minimum value of
350 km. Most of the BPs always exist during the observational time
interval, and this means that their lifetimes are at least tens of
minutes.

\begin{figure*}
   \centering
   \sidecaption
   \includegraphics[bb=50 240 510 587,width=12cm]{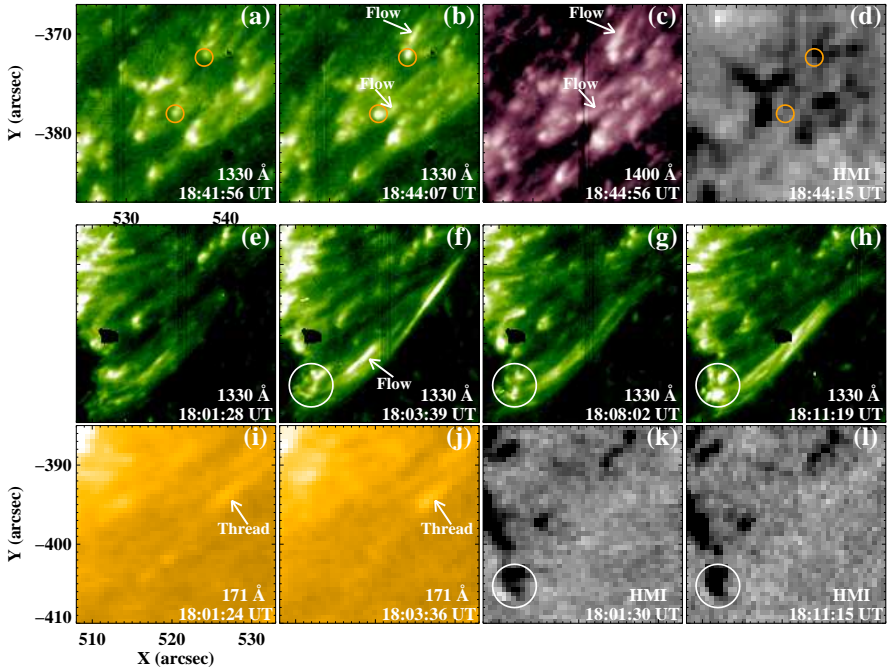}
   \caption{Panels
(a)-(d): temporal evolution of two BPs (yellow circles), associated
upflows (white arrows) in IRIS 1330 {\AA} and 1400 {\AA} images and
the locations of BPs in HMI LOS magnetogram. Panels (e)-(h): IRIS
1330 {\AA} images showing the repeated appearance of another drastic
upflow. Panels (i)-(l): SDO/AIA 171 {\AA} images showing the
relatively steady EUV threads, and HMI magnetograms presenting
magnetic flux cancellation at the concentration area of several
BPs.}
   \label{F:lineprofiles}%
\end{figure*}

The BPs seem to be connected with loop-like structures, appearing as
the morphology of ``tadpoles". At region ``R1", the directions of
the loop-like structures are roughly parallel to the spine of the
filament, pointing from the BPs to the northwest (Figures 2b and
2e). The comparison of the IRIS and AIA images suggests that several
loop-like structures at 1330 and 1400 {\AA} could be observed at 171
and 304 {\AA}, corresponding to the bright EUV threads of the
filament (arrows in Figures 2a-b and 2d-e). The BPs are located at
the negative-polarity fields (Figure 2c) near the polarity inversion
line (PIL). Several relatively isolated BPs are defined and their
magnetic flux (see white circles in Figures 2b-c) is calculated. The
magnetic flux of these BPs is between 3.8$\times$10$^{17}$ Mx and
1.2$\times$10$^{18}$ Mx, and the average value is about
7.0$\times$10$^{17}$ Mx. Signatures of these BPs and loop-like
structures are not very prominent in the chromospheric 2796 {\AA}
images (Figure 2f).

\begin{figure*}
   \centering
   \sidecaption
   \includegraphics[bb=54 309 510 536,width=12cm]{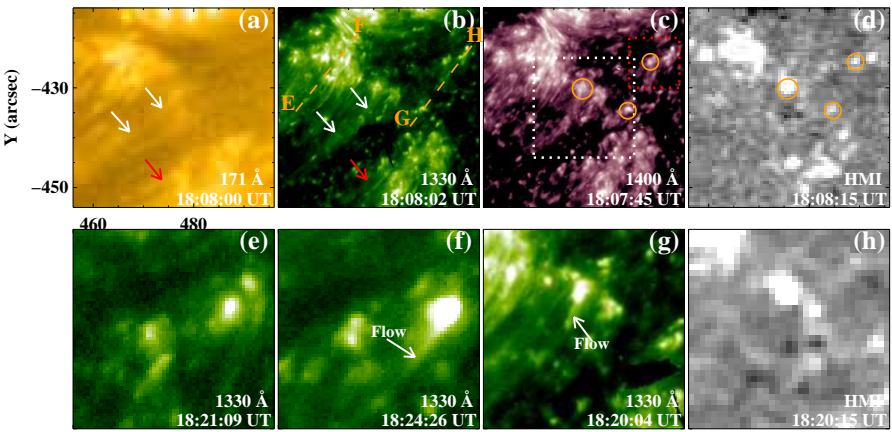}
   \caption{Appearance of BPs and upflows at region ``R2" (white square
in Figure 1c). The arrows (panels a-b) point to the loop-like
structures connecting with BPs in AIA 171 {\AA} and IRIS 1330 {\AA}
images. Dashed lines ``E$-$F" and ``G$-$H" (panel b) show the cut
positions used to obtain the stack plots shown in Figures 5(c)-(d).
The yellow circles outline the locations of several BPs in the 1400
{\AA} image and HMI LOS magnetogram. Red and white squares in panel
(c) respectively denote the FOVs of panels (e)-(f) and (g)-(h).}
   \label{F:timeslices}%
\end{figure*}

\begin{figure*}
   \centering
   \sidecaption
   \includegraphics[bb=68 270 473 579,width=12cm]{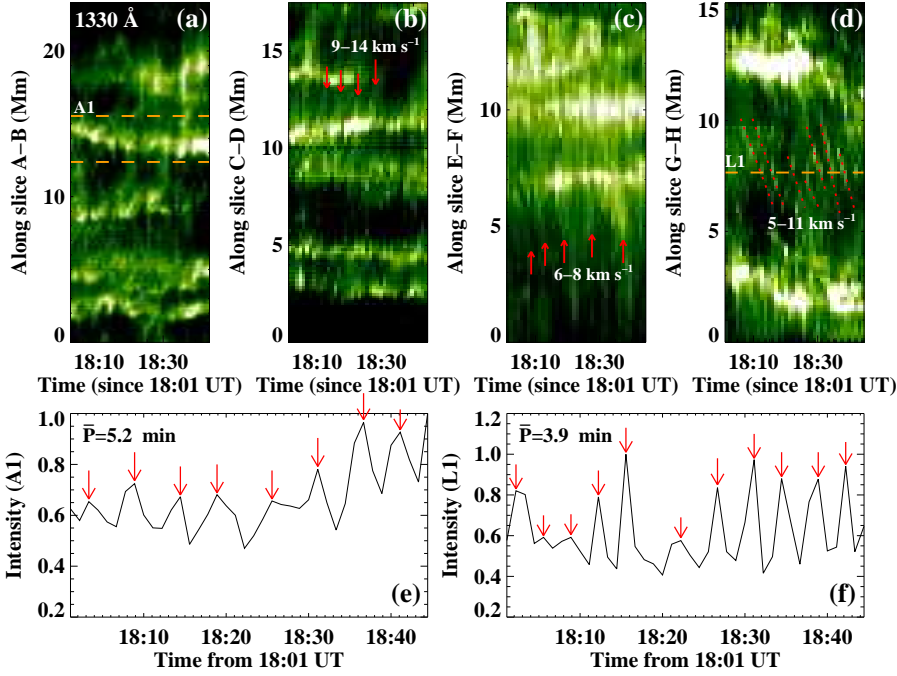}
   \caption{Panels
(a)-(d): stack plots along slices ``A$-$B" and ``C$-$D" (dashed
lines in Figure 2b) and slices ``E$-$F" and ``G$-$H" (dashed lines
in Figure 4b). Red arrows and dotted lines show the repeated
upflows. Panel (e): intensity variation of a BP in panel (a) at 1330
{\AA}. Each data point of the profile is the vertical sum of the
intensity in area ``A1" (two dashed lines in panel a). Red arrows
point to the peaks of the quasi-periodic intensity profile. Panel
(f): horizontal slice (black curve) along the dashed line (``L1") in
panel (d) at 1330 {\AA}.}
   \label{F:spectrographs2}%
\end{figure*}

\begin{figure}
   \centering
   \includegraphics[bb=55 187 484 631,width=8.8cm]{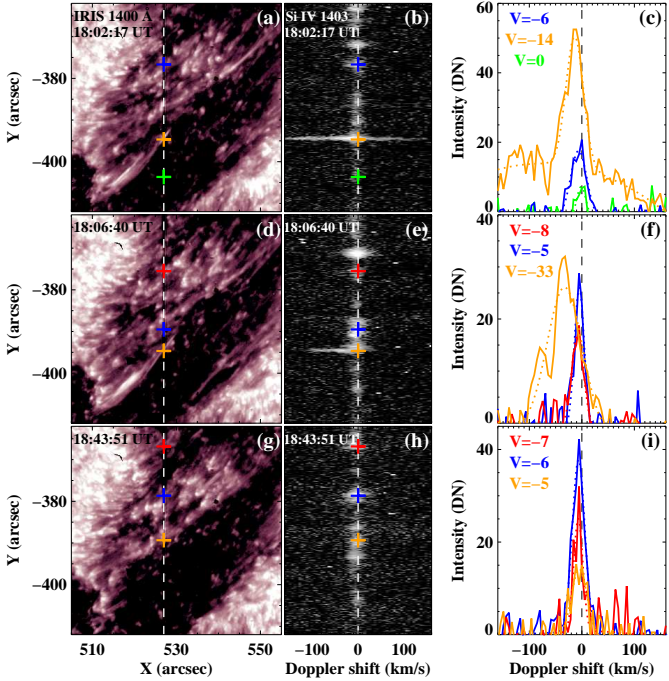}
   \caption{IRIS 1400 {\AA} images (left column), Si IV
1402.77 {\AA} spectra (middle column) and profiles of the Si IV line
(right column) at the selected locations. The FOV of 1400 {\AA}
images is the same as that of Figure 2. The red, blue and yellow
pluses denote the locations of the upflows initiated from the BPs.
The green plus in panel (a) indicates one location without upflows.
Dotted curves in right column are the single Gaussian fitting with a
parabolic background of the solid curves.}
   \label{F:spectra2}%
\end{figure}

The temporal evolution of some BPs at region ``R1" is presented in
Figure 3. Red and white squares in Figure 2(e) respectively denote
the FOVs of Figures 3(a)-(d) and 3(e)-(l). We find that the emission
intensities of the BPs at 1330 {\AA} vary with time. At 18:41:56 UT,
the intensities of the two BPs (yellow circles in Figure 3a) were
very weak. About 2 min later, the emissions of the two BPs increased
by 60\% and the mass upflows were simultaneously initiated (Figures
3b-c). The constant upflows resulted in the appearance of observed
loop-like structures linked with the BPs. The magnetic fields at the
two BPs are very weak and could not be clearly discerned (Figure
3d). More drastic events occasionally occurred at the concentration
area of the BPs (Figures 3e-h). At about 18:03 UT, the emission
intensity of the BPs was enhanced and meanwhile the upflow occurred
with a projected velocity of about 50 km s$^{-1}$. Then the
intensity of the upflow was decreased at 18:08 UT. About 3 min
later, the BPs and the upflow were activated again. In AIA 171 {\AA}
images (panels i-j), the outflows (e.g., see the ``Flow" in panel
(f)) could not be observed. Only the relatively steady EUV threads
could be discerned. At the base of the upflow, magnetic flux
cancellation of the strong negative polarity fields with the nearby
small-scale positive polarity fields was observed (Figures 3k-l).

The appearance of the BPs at region ``R2" (Figure 1c) was shown in
Figure 4. The loop-like structures connecting with the BPs could be
observed in the 171 and 1330 {\AA} channels, corresponding to the
bright EUV threads of the quiescent filament (arrows in Figures
4a-b). The direction of the loop-like structures at this region is
roughly towards the southeast, which is different from the loop-like
structures at region ``R1" (red and white arrows in Figure 1c). The
BPs are anchored at positive polarity fields and their magnetic flux
ranges from 6.9$\times$10$^{17}$ Mx to 1.4$\times$10$^{18}$ Mx
(Figures 4c-d). The small-scale explosive events are frequently
initiated in close proximity to some BPs. Taking one BP (Figures
4e-f) as an example, its emission intensity at 1330 {\AA} was
suddenly enhanced and meanwhile the bright material was ejected from
the BP with a velocity of 10 km s$^{-1}$. The BP and the associated
upflow comprise a ``tadpole-like" morphology. At 18:20:04 UT, the
upflow at another BP appeared (Figure 4g). The magnetogram showed
that extremely weak negative polarity fields seemed to exist nearby
the main positive polarity fields (Figure 4h).

In order to analyze the kinematic evolution of the upflows at 1330
{\AA} in detail, we obtain the stack plots (Figures 5a-d) along four
different straight lines (``A$-$B" and ``C$-$D" in Figure 2b;
``E$-$F" and ``G$-$H" in Figure 4b). The stack plots show multiple
moving intensity features, with each strip representing the
propagating upflow. The velocity of the upflows initiated from the
BPs was about 5$-$14 km s$^{-1}$. The emission of the BPs and the
appearance of upflows are intermittent, and indeed somewhat
periodic. The upflows seemed to appear each time when the emission
intensity of the BP was enhanced. The vertical sum of the intensity
in area ``A1" (two dashed lines in Figure 5a) was shown in Figure
5(e). The intensity of the BP exhibited a quasi-periodic pattern,
with an average period of about 5.2 min. The peak intensity of the
BP was about 1.4 times as high as the valley intensity. The upflows
associated with the brightenings of BPs also occurred periodically.
In the stack plot at region ``R2" (Figure 5d), we select the
intensity variation at location ``L1" to obtain the period of the
upflows. Location ``L1" is far away from the north BPs and thus the
intensity variation of the BPs is not affecting the measured
intensity of the upflows. Also, several upflows could be clearly
detected at this location. It shows that the period of the upflows
was about 3.9 min (black curve in Figure 5f).

Figure 6 shows the profiles of Si IV 1402.77 {\AA} line at locations
of the upflows from the BPs. We are interested mostly in the main
component of the line, including the line shift and the line width.
Therefore we add a parabolic background in the single Gaussian fit
just to account for the wing emission. The FWHW of the fitting
profile is thought to be the line width (Peter et al. 2014). The
recurrent upflows generated from the same BP have different Doppler
shifts and line widths. At 18:02:17 UT, the emission profile at the
notable upflow (yellow plus in Figure 6a) was blueshifted by
$\sim$14 km s$^{-1}$, with the Doppler width of about 40 km s$^{-1}$
(Figures 6b-c). For the use of a single Gaussian function with a
parabolic background, the fitting profile at this location (yellow
dotted line in panel c) looks asymmetric and shows a blue-wing
excess. One can have used also a Gaussian function for the extended
wings (Peter 2000), but in the present case the parabolic fit works
as well. The inclusion of a second broad component can also have an
impact on the shift of the main component. The broad component that
we account for by the parabolic background could be, i.e. that some
further high-velocity component is present (indicating jets or
whatever; see Peter 2001, 2010), but this contributes less to the
emission than the main component of the spectrum. The emission
intensity of the second upflow at the same location was decreased by
40\% at 18:06:40 UT (yellow pluses and curves in Figures 6d-f).
However, the blueshift of the second upflow reached about 33 km
s$^{-1}$ and the line width was increased by a factor of two.
Afterwards, the upflows exhibited small blueshifts of 5$-$7 km
s$^{-1}$ and had the line width of about 20$-$30 km s$^{-1}$
(Figures 6g-i).

\section{Summary and Discussion}

We firstly report the observations of subarcsecond BPs underlying a
quiescent filament in the TR by the IRIS on 2015 February 13. The
BPs have a spatial scale of about 350$-$580 km and lifetime of more
than several tens of minutes. The dimension of BPs is similar to
those of the EUV bright dots at the edge of the active region
(R{\'e}gnier et al. 2014) and the subarcsecond bright dots in the TR
above sunspots (Tian et al. 2014a), however, the lifetime of BPs is
much longer than those of the reported bright dots ($\sim$25 s).
Compared to well-known CBPs, the BPs in the filament channel are
much smaller than CBPs (with a spatial size of 5$-$20$\arcsec$;
Vaiana et al. 1973). The subarcsecond BPs and CBPs are both
long-lived and their emission intensities both exhibit the
quasi-periodic variation. In this work, the intensities of BPs have
a period of about 4$-$5 min, comparable to the minimum period of
CBPs. The BPs are anchored at small-scale magnetic structures near
the PIL and their magnetic flux was about 10$^{17}$$-$10$^{18}$ Mx,
almost two orders of magnitude smaller than that of CBPs.

The directions of upflows are almost parallel to the spine of the
quiescent filament. The upflows initiated from the BPs with opposite
polarity magnetic fields have opposite directions (Figure 1c). The
direction of the upflows is generally thought to trace out the
connectivity of the magnetic fields. The bi-directional upflows
probably imply the consistent direction of magnetic fields along the
filament spine. The filament is composed of multiple groups of
magnetic systems and their directions are the same, pointing from
the northwest to the southeast (Figure 1). The connectivity between
magnetic structures within the filament channel is not in any
arbitrary pattern, e.g., the positive magnetic structures in ``R2"
connect with the southeast negative ones and the negative magnetic
structures in ``R1" connect with the northwest positive ones
(Figures 1c and 1e). Martin (1998) found that the chromospheric
fibrils on the two sides of an H$\alpha$ filament had opposite
orientations. Sheeley et al. (2013) reported cellular plumes leaning
in opposite directions on the two sides of a filament channel based
on SDO observations. The configuration of the upflows within the
filament channel in our work is roughly consistent with those shown
in previous studies and may be an intrinsic feature of the quiescent
filament. The multiple groups of magnetic systems along the filament
spine have the same orientation and this magnetic topology is more
stable than the opposite field orientations within the filament due
to the low probability of magnetic reconnection.

The BPs appear to be the bases of filament threads and the upflows
intermittently inject plasma inside the fine threads. Our
observations imply that the source of the filament material may be
the recurrent upflows in the TR. Chae (2003) and Liu et al. (2005)
reported the findings of H$\alpha$ jets associated with the
filaments and suggested that the frequent injection of flow from
below supplies the mass necessary for the formation of the
filaments. In our work, the quasi-periodic upflows might be able to
convey mass into the filament threads, which is consistent with the
injection models that the filament plasma is lifted from the lower
atmosphere by magnetic reconnection (Wang 1999; Litvinenko et al.
2007).

Quasi-periodic upflows are generated when the intensities of BPs are
enhanced, with the BPs and jet-like upflows composing a morphology
of ``tadpoles". The quasi-periodic brightenings of BPs and
appearances of upflows might be caused by oscillatory magnetic
reconnection (Kankelborg et al. 1997; Gupta \& Tripathi 2015). The
recurring time scale of the upflows is about 4$-$5 min,
approximately consistent with the period of solar p-mode waves. This
implies that p-mode waves possibly modulate magnetic reconnections
by changing the plasma density near the reconnection site and induce
periodic reconnection of the magnetic fields (Chen \& Priest 2006).
The brightenings of BPs and generation of upflows are observed each
time when the magnetic reconnection occurs. Recently, Li \& Zhang
(2015) found quasi-periodic oscillations of the small-scale
substructures within the flare ribbon with periods of 3-6 min, which
is attributed to quasi-periodic slipping reconnection. The
observation of wave-like pulses in a filament pillar was presented
and simulated by Schmieder et al. (2013) and Ofman et al. (2015).
These waves have a period of about 300 s, similar to that of
quasi-periodic upflows in our event. They interpreted these waves
propagating perpendicular to the filament magnetic field as
nonlinear fast magnetosonic waves in the filament foot. In this
work, the velocity of the upflows in plane of sky is about 5$-$50 km
s$^{-1}$. The emission line of Si IV 1402.77 {\AA} at the locations
of upflows exhibits obvious blueshifts of about 5$-$30 km s$^{-1}$,
and the line profile is broadened with the width of more than 20 km
s$^{-1}$. The Doppler velocities of the repeated upflows at the same
location are not constant and change with time. This may be caused
by the different energy released during the magnetic reconnection
process. The more energy is released, the higher velocity the upflow
has.

High-resolution observations showed that filament threads had an
average width of 0$\arcsec$.3 (Lin et al. 2005; Li \& Zhang 2013),
slightly thinner than the spatial scale of subarcsecond BPs
(0$\arcsec$.48$-$0$\arcsec$.8). The plasma in the filament threads
moves with the magnetic field lines in opposite directions,
appearing as ``counter-streaming" flows (Zirker et al. 1998;
Schmieder et al. 2010; Alexander et al. 2013). The speed of the
``counter-streaming" flows is about 5$-$20 km s$^{-1}$, comparable
to that of quasi-periodic upflows at the bases of filament threads.
Until now, the mass source and triggering mechanism of anti-parallel
flows are still unknown. Chen et al. (2014) suggested that the
H$\alpha$ counter-streamings in previous observations were caused by
longitudinal oscillations of filament threads. Our observations
reveal that the upflows at the BPs with opposite polarity magnetic
fields move along the filament threads in opposite directions. The
bi-directional upflows travelling from the positive and
negative-polarity footpoints of filament threads indicate that the
filament is composed of multiple separate systems of magnetic
fields. The ``counter-streaming" flows may be caused by the
propagation of bi-directional upflows along different magnetic
systems that are initiated at the subarcsecond BPs below the
filament.

\begin{acknowledgements}

IRIS is a NASA small explorer mission developed and operated by
LMSAL with mission operations executed at NASA Ames Research center
and major contributions to downlink communications funded by the
Norwegian Space Center (NSC, Norway) through an ESA PRODEX contract.
We acknowledge the SDO/AIA and HMI for providing data. This work is
supported by the National Natural Science Foundations of China
(11303050, 11533008 and 1221063) and the Strategic Priority Research
Program$-$The Emergence of Cosmological Structures of the Chinese
Academy of Sciences, Grant No. XDB09000000.

\end{acknowledgements}

\end{document}